\documentclass[conference]{IEEEtran}
\IEEEoverridecommandlockouts
\usepackage{cite}
\usepackage{amsmath,amssymb,amsfonts}
\usepackage{algorithmic}
\usepackage{graphicx}
\usepackage{textcomp}
\usepackage{xcolor}
\usepackage{subcaption}
\usepackage{braket}
\usepackage{qcircuit}
\usepackage{amsmath}
\usepackage{array}
\usepackage{titlesec}
\usepackage{makecell}
\usepackage{tabularx}
\usepackage{siunitx}
\newcolumntype{?}{!{\vrule width 1.5pt}}
\newcommand{\ceil}[1]{\left\lceil #1 \right\rceil}
\usepackage[hidelinks]{hyperref}
\def\BibTeX{{\rm B\kern-.05em{\sc i\kern-.025em b}\kern-.08em
    T\kern-.1667em\lower.7ex\hbox{E}\kern-.125emX}}
\begin{document}

\title{AMARETTO: Enabling Efficient Quantum Algorithm Emulation on Low-Tier FPGAs \vspace{-10pt}
\thanks{ \vspace{-3pt}This work was supported in part by AMD under the  \href{https://www.amd.com/en/corporate/university-program.html}{AMD University program} }
}

\author{\IEEEauthorblockN{Christian Conti\IEEEauthorrefmark{1}, Deborah Volpe\IEEEauthorrefmark{1}, Mariagrazia Graziano\IEEEauthorrefmark{1},  Maurizio Zamboni\IEEEauthorrefmark{1}, and Giovanna Turvani\IEEEauthorrefmark{1} }
\IEEEauthorblockA{\IEEEauthorrefmark{1}Politecnico di Torino
Italy\\
\{christian.conti, deborah.volpe, mariagrazia.graziano, maurizio.zamboni, giovanna.turvani\}@polito.it} \vspace{-30pt}}

\maketitle

\begin{abstract}
Researchers and industries are increasingly drawn to quantum computing for its computational potential. However, validating new quantum algorithms is challenging due to the limitations of current quantum devices. Software simulators are time and memory-consuming, making hardware emulators an attractive alternative.\\
This article introduces AMARETTO (quAntuM ARchitecture EmulaTion TechnOlogy), designed for quantum computing emulation on low-tier Field-Programmable gate arrays (FPGAs), supporting Clifford+T and rotational gate sets. It simplifies and accelerates the verification of quantum algorithms using a Reduced-Instruction-Set-Computer (RISC)-like structure and efficient handling of sparse quantum gates. A dedicated compiler translates OpenQASM 2.0 into RISC-like instructions. AMARETTO is validated against the Qiskit simulators. Our results show successful emulation of sixteen qubits on a AMD Kria KV260 SoM. This approach rivals other works in emulated qubit capacity on a smaller, more affordable FPGA.
\end{abstract}

\vspace{-6pt}
\section{Introduction}
\vspace{-3pt}
\noindent In recent years, interest in \textbf{quantum computing} has achieved unique acceleration thanks to its potential in data-intensive applications. Nonetheless, the \textbf{validation} of new quantum computing algorithms is challenging due to the constraints imposed by current quantum devices. The production, administration, and upkeep of quantum hardware are exclusive domains of major corporations that grant access through cloud-based platforms, although usually with fees. Furthermore, the fidelity of outcomes can be substantially compromised by devices' noise.\\ \textbf{Classical simulation} remains the most popular solution for debugging, providing insights into the quantum state, hard to retrieve on real quantum hardware, but \textbf{software simulation} faces drawbacks, such as long execution times and high memory requirements, limiting scalability. Hence, exploring \textbf{classical hardware platforms} such as \textbf{Field-Programmable Gate Arrays (FPGAs)} holds significant promise. Indeed, hardware emulators are expected to outperform software-based counterparts in simulating quantum phenomena due to their ability to replicate the parallel nature of quantum computation more accurately.
\\
This work introduces \textbf{AMARETTO} (\textbf{quAntuM ARchitecture EmulaTion TechnOlogy}), a \textbf{Reduced-Instruction-Set-Computer (RISC)}-like architecture for quantum emulation on \textbf{low-tier FPGAs}, supporting \textbf{Clifford+T} and \textbf{rotational gate sets}. Validated using the \href{https://qiskit.github.io/qiskit-aer/apidocs/aer_provider.html}{Qiskit simulators}, AMARETTO successfully emulated \textbf{sixteen qubits} on the \textbf{AMD Kria KV260 SoM} using a \textbf{twenty-bit fixed-point} numeric representation. This approach matches other works' qubit capacity but with a smaller, more accessible FPGA.\\
The article's organization includes a review of quantum simulation on classical platforms and related work (\textbf{Section \ref{sec:Background}}), details of the proposed architecture (\textbf{Section \ref{sec:AMARETTO}}), results and validation methodology (\textbf{Section \ref{sec:results}}), and conclusions with future perspectives (\textbf{Section \ref{sec:conclusions}}).
\vspace{-6pt}
\section{Background and related works}\label{sec:Background}
\vspace{-5pt}
\subsection{Quantum computing emulation}
\begin{figure}[h]
    \centering
            \scalebox{0.7}{
        \Qcircuit @C=7em @R=0.1em @!R {
        \lstick{\mbox{\textbf{$\ket{\psi_{t_0}}$}}} &  \rstick{\mbox{\textbf{$\ket{\psi_{t_1}} = \textrm{G}_1 \ket{\psi_{t_0}}$}}} &  \rstick{\mbox{\textbf{$\ket{\psi_{t_2}} = \textrm{G}_2 \ket{\psi_{t_1}}$}}} &\\
        \lstick{ {q}_{0} : \ket{0}  } \barrier[-2em]{1} &\gate{H} \barrier{1} & \ctrl{1} \barrier[-2em]{1}   & \meter  \\
        \lstick{ {q}_{1} : \ket{0}  } & \qw & \targ    & \meter  \\
    & \mbox{\textbf{$\textrm{G}_1$}} &  \mbox{\textbf{$\textrm{G}_2$}}  & \\ 
        }}
        \centering
  \caption{Example of a two-qubit quantum circuit, highlighting with dotted vertical lines the different layers and showing on the top the state vector evolution layer by layer. \vspace{-15pt} }
	    \label{fig:QuantumCircuit}
\end{figure}
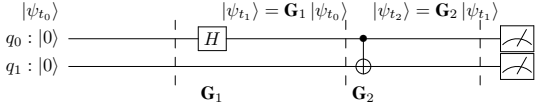

\vspace{-3pt}
Quantum computing is a new computational paradigm, leveraging quantum mechanic principles like \textbf{superposition} and \textbf{entanglement}. Its fundamental unit, the \textbf{qubit}, can exist in \textbf{infinite possible states}, unlike a classical bit which is either 0 or 1. Using Dirac notation, a qubit's state is expressed as the \textbf{state vector}:
\vspace{-3pt}
\begin{equation}
    \ket{\psi} = a \ket{0} + b \ket{1} = a \begin{pmatrix}1\\0\end{pmatrix} + b\begin{pmatrix}0\\1\end{pmatrix} = \begin{pmatrix}a\\b\end{pmatrix} \, ,
    \vspace{-3pt}
    \label{eq:oneQubitState}
\end{equation}
where $\ket{0}$ and $\ket{1}$ are the basis states and $a$ and $b$ are complex \textbf{probability amplitudes}. When measured, the qubit collapses to $\ket{0}$ or $\ket{1}$ with probabilities $|a|^2$ and $|b|^2$, respectively.\\
An $n$-qubit system's state is represented by the tensor product of individual qubit states:
\vspace{-3pt}
\begin{align}
\begin{split}
    &\ket{\psi} = \ket{\psi_{n-1}}  \otimes \ket{\psi_{n-2}}  \otimes \cdots \otimes  \ket{\psi_{1}} \otimes \ket{\psi_{0}} =  \vspace{-15pt}\\
  &= c_{0 \cdots0} \ket{0 \cdots0} +  \cdots  + c_{1 \cdots1} \ket{1 \cdots1} \, ,
\end{split}
 \vspace{-3pt}
\label{eq:MultiQubitState}
\end{align}
\textbf{Quantum gates}, described by \textbf{unitary matrices} of dimension $2^m \times 2^m$, where $m$ is the number of involved qubits, modify the system state. Gates involving multiple qubits can create \textbf{entanglement}, leading to strong correlations between qubits.\\
To classically simulate a \textbf{quantum circuit}, which entails a series of transformations,  it is necessary to compute the product of a $2^n \times 2^n$ matrix and the $2^n$ state vector for each gate layer. These matrices arise from the tensor product of gate matrices for each qubit, assuming the identity matrix when no gate targets a specific qubit, as illustrated in Figure \ref{fig:QuantumCircuit}. Therefore, the \textbf{scalability} challenges due to the \textbf{exponential increase in complexity} with the qubits count affecting \textbf{operations} and \textbf{memory}  become evident.

For more details about quantum computing, refer to \cite{nielsen_quantum_2010}.

\vspace{-0pt}

\vspace{-3pt}
\subsection{Previous work}
\vspace{-3pt}
In recent years, various FPGA architectures have emerged to tackle the limitations of software emulation in quantum computing. For instance, \cite{pilch2019fpga} introduced an emulator that loads layer matrices from a processor and computes a parallel product with the state vector to determine the new state. However, scalability becomes challenging due to the exponential increase in computation and memory demands with qubit count growth. \cite{mahmud2018scalable} also employed parallel matrix-vector product technique, enhancing precision by using floating-point number representation but with constraints on the number of emulated qubits. In another work \cite{mahmud2020efficient}, scalability was enhanced by moving the storage of the state vector from the FPGA to external memory.\\
The approach described in \cite{fujishima200316} emulates quantum circuits by computing interactions among basis states on an $N$-dimensional hypercube, reaching the emulation of sixteen qubits.
\begin{figure}[h]
    \centering \vspace{-15pt}
    \includegraphics[width=1\columnwidth]{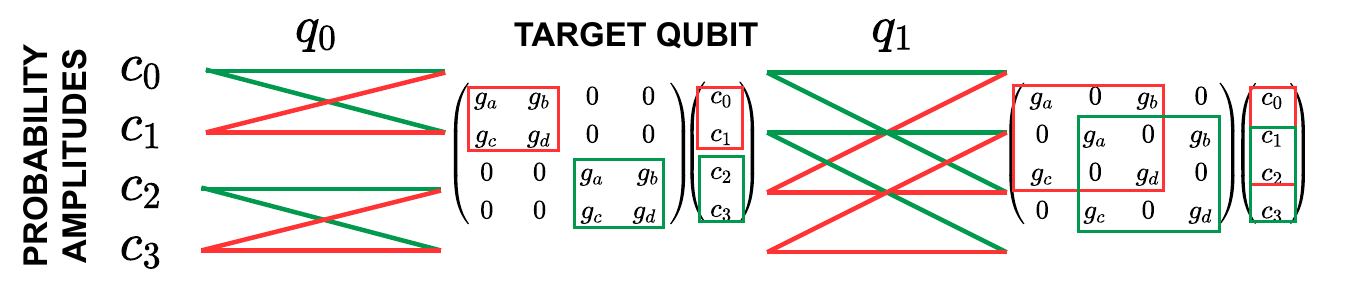}\vspace{-3pt}
    \caption{The butterfly-like mechanism for selecting interacting couples of probability amplitudes in a two-qubit system. \vspace{-6pt}}
    \label{fig:Butterfly}
\end{figure}
\\
In \cite{conceiccao2015efficient}, as in AMARETTO, a \textbf{butterfly-like} selection mechanism for interacting couples in the state vector was utilized to reduce unnecessary operations. However, this architecture's area requirement significantly increases with the qubit count and supports only a limited set of gates (Pauli X, CNOT, Toffoli, and Hadamard), restricting its applications. In contrast, AMARETTO supports a universal quantum gate set, enabling the execution of any circuit type.

\vspace{-3pt}
\section{AMARETTO: an efficient quantum emulator}\label{sec:AMARETTO}

\begin{figure*}[t]
    \centering
    \begin{subfigure}[t]{1.1\columnwidth}
        \centering
        \includegraphics[width=1\textwidth]{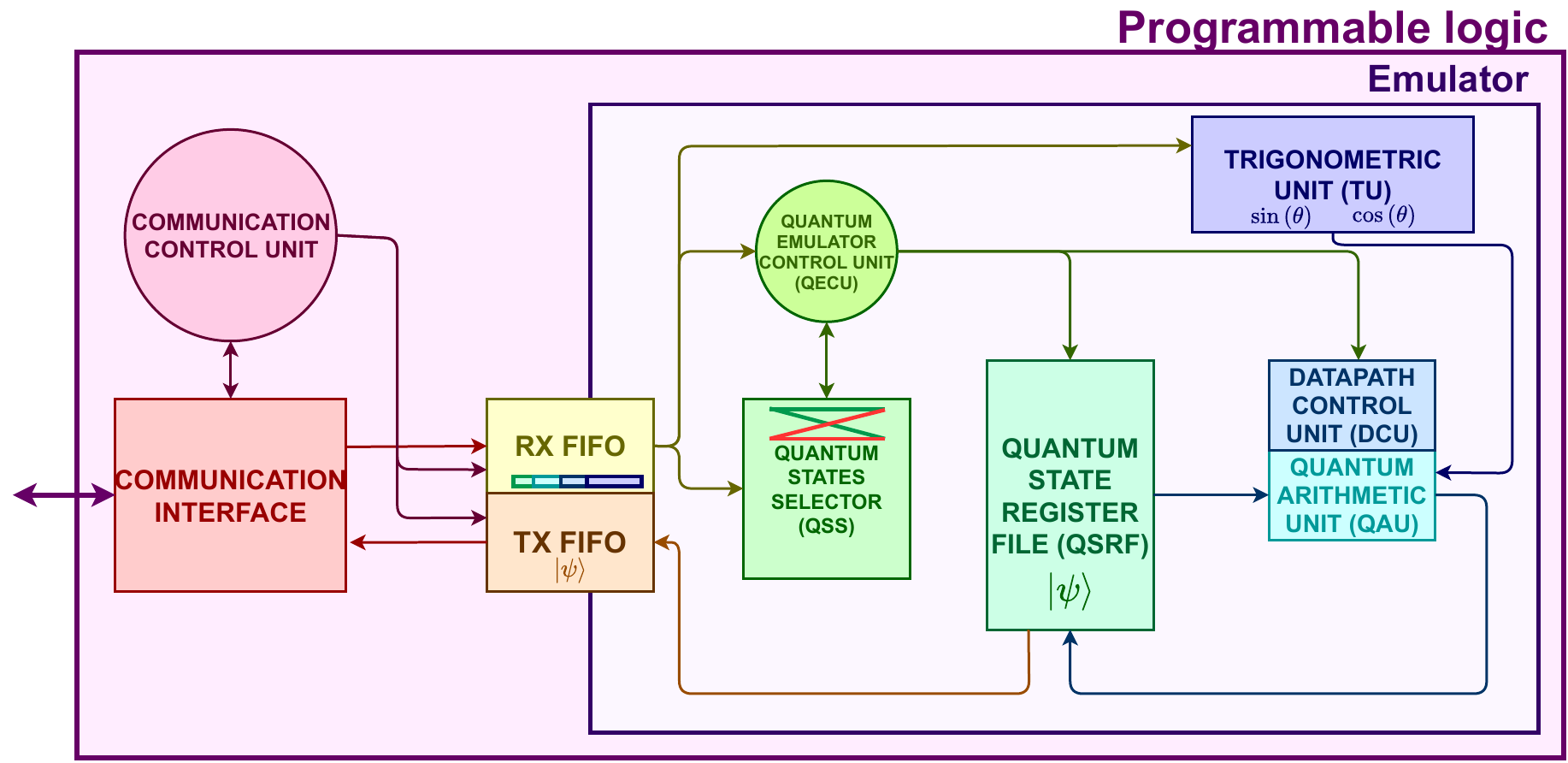}
        \caption{AMARETTO architecture: comprising a register for state vector elements, a state selector executing the butterfly algorithm, a computing unit, a Trigonometric Unit (TU), and a central control unit (QECU).}
        \label{fig:AMARETTO_RISC_like_structure}
    \end{subfigure}
    \quad\quad
    \begin{subfigure}[t]{0.7\columnwidth}
        \centering
        \includegraphics[width=1\textwidth]{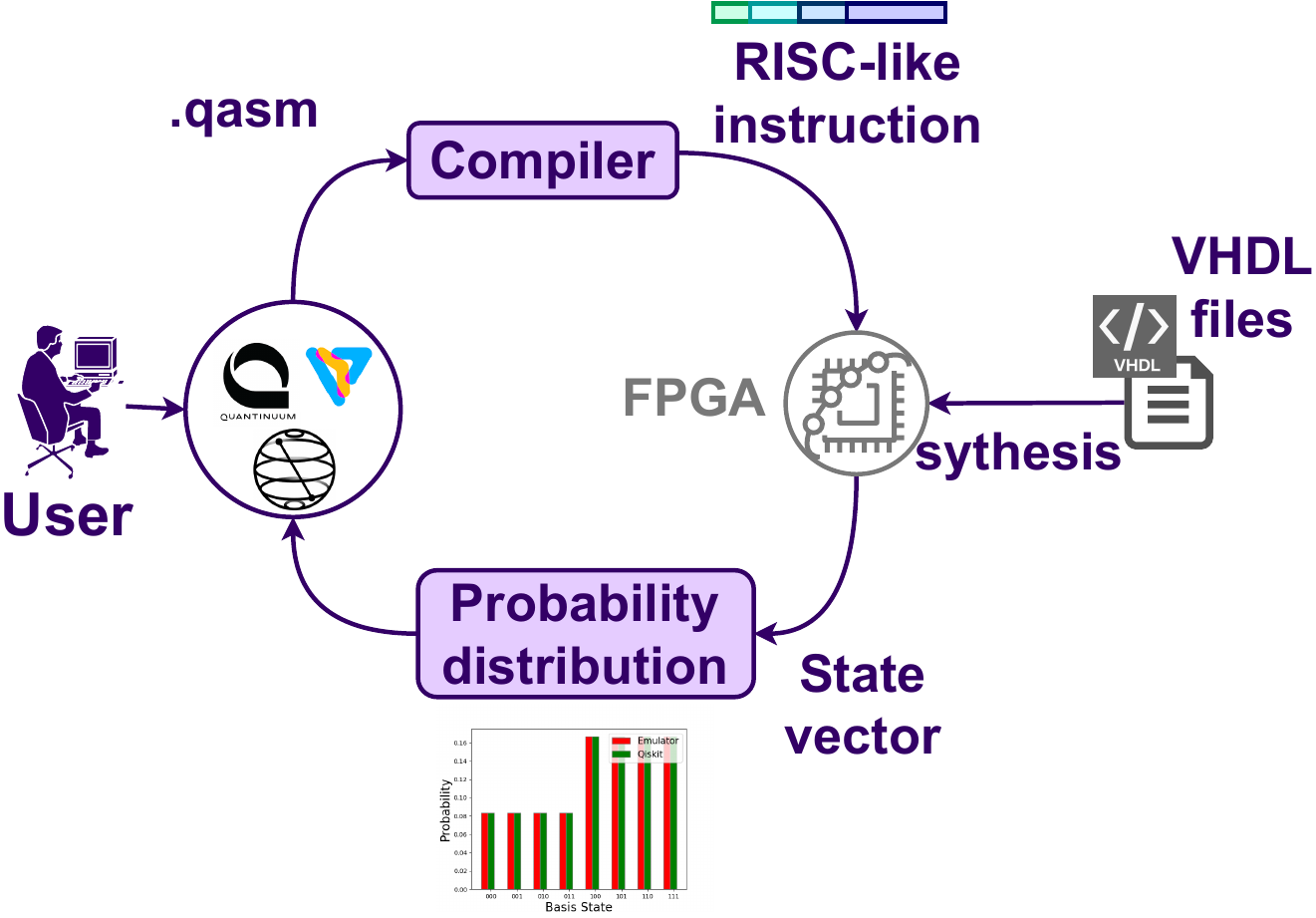}
        \caption{High-level description of the AMARETTO emulation environment.}
        \label{fig:AMARETTO_emulator}
    \end{subfigure}
    \caption{AMARETTO architecture and high-level scheme of its emulation environment. \vspace{-15pt}}
    \label{fig:AMARETTO_scheme}
\end{figure*}

\textbf{AMARETTO} (\textbf{quAntuM ARchitecture EmulaTion TechnOlogy}) is an efficient architecture for \textbf{quantum computing emulation on FPGA} platforms, supporting \textbf{Clifford+T and rotational gate sets}, and designed in VHDL for implementation within any modern FPGA by leveraging embedded blocks like Random Access Memories (RAMs) and Digital Signal Processing (DSP) blocks. This \textbf{portability} across devices is achieved by modifying the \textbf{communication interface} (Figure \ref{fig:AMARETTO_RISC_like_structure}).\\
The architecture strategically reduces computational complexity by employing a \textbf{butterfly-like} mechanism, as detailed in Section \ref{sec:Background} and shown in Figure \ref{fig:Butterfly}. This approach isolates interacting probability amplitudes essential for obtaining the output state vector, capitalizing on the sparse nature of equivalent gate matrices. In this way, it is possible to avoid non-operations and the computation of the equivalent layer gate matrix. Two-qubit controlled gates can be implemented by filtering interacting couples associated with the basis state where the control qubit is equal to one. Moreover, a \textbf{20-bit fixed-point} number representation (2 bits for decimal and 18 for fractional parts) with a \textbf{nearest-even} approximation mechanism is chosen, allowing a reduction of both the area and complexity of arithmetic operators with respect to the floating point one. This also reduces the memory requirements for saving a probability amplitude, leading to more emulable qubits on the same platform. The precision of number representation was chosen based on analysis conducted for the butterfly-based mechanism in \cite{LagostinaThesis}, which proves that the accuracy of the simulation is not significantly affected by the approximation. 
\\
The AMARETTO environment prioritizes user-friendliness (Figure \ref{fig:AMARETTO_emulator}), allowing potential users to describe the quantum circuit using leading quantum frameworks. These generate \textbf{OpenQASM 2.0} \cite{cross2017open}, which is then processed by the \textbf{compiler} to translate the gates into a set of supported instructions transmitted to the emulator. Upon completing the simulation process, the user receives the probability amplitudes of the final state vector, providing data ready for user analysis.
\begin{figure}[h]
    \centering \vspace{-13pt}
    \includegraphics[width=0.8\columnwidth]{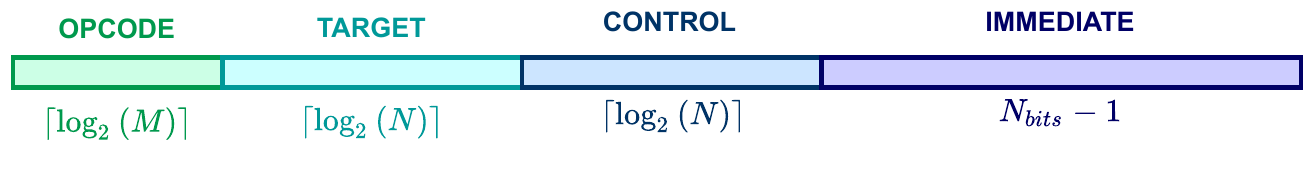}\vspace{-5pt}
    \caption{AMARETTO g-type instruction, separating the fields. \vspace{-9pt}}
   \label{fig:OperationsFormat}
\end{figure}
\\
The instructions can be classified into three types: the \textbf{s-type} for setting the number of qubits in the circuit, the \textbf{g-type} for executing gates, and the \textbf{r-type} for reading the state vector. As shown in Figure \ref{fig:OperationsFormat}, the g-type instructions include an opcode identifying the gate, bits defining the target and control qubits --- in case a single-qubit gate, target and control field coincide ---, and an immediate field containing normalized angles. For s-type, the immediate is equal to the number of qubits, while for r-type, only the opcode field is relevant. 
For the target FPGA, an instruction is 32-bit long since five bits are considered for the opcode, eight bits are for target and control qubits identification (sixteen emulable qubits), and nineteen bits are associated with immediate (an angle represented in the range $[-1,1)$, with eighteen bits for the fractional part). However, the length of the instruction varies based on both the number of emulable qubits on the target FPGA and the number of fractional bits considered for precision. Furthermore, the compiler indirectly supports additional gates --- in particular, all the gates supported by the OpenQASM 2.0 --- by leveraging known equivalences in the literature.\\
AMARETTO follows a \textbf{RISC-like} structure (Figure \ref{fig:AMARETTO_RISC_like_structure}), including a \textbf{register file}, which stores the real and imaginary parts of the state vector (\textbf{Quantum State Register File}, \textbf{QSRF}), the data path for evaluating gate effects on probability amplitude interacting couples (\textbf{Quantum Arithmetic Unit}, \textbf{QAU}), a \textbf{Quantum State Selector} (\textbf{QSS}),  implementing the butterfly selection mechanism, a \textbf{Trigonometric Unit} (\textbf{TU}), which computes sine and cosine, a \textbf{control unit} (Quantum Emulator Control Unit, \textbf{QECU}) and the \textbf{communication interface} responsible for receiving architecture instructions and managing the state vector. \\
The \textbf{QSRF}, sized at $2^N$ elements (where $N$ is the number of qubits), optimizes space utilization by taking full advantage of the BRAM blocks, operating with two output and one input ports clocked at double the nominal frequency. This configuration,  called \textbf{pumping} \cite{pumping}, enables the reading and writing of two probability amplitudes in each clock cycle.  \\
Differently from the previous works, AMARETTO computes  \textbf{couple by couple the probability amplitudes} to minimize area requirements and increase the number of simulable qubits. The \textbf{instruction level parallelism} can be exploited by introducing five pipeline levels. This strategic implementation significantly reduces time penalties, taking advantage of the absence of data dependencies in the execution of a single gate since the interacting couples are independent of each other. The pipeline reaches its maximum effectiveness when the number of couples to update is equal to or higher than the number of pipeline stages, i.e., when $2^{N_q-2} \geq N_\textrm{pipe} \rightarrow N_q \geq \ceil{\log_2{(N_\textrm{pipe})}+2} = N_{q_\textrm{min}}$, where $N_q$ represents the number of qubits in the circuit and $N_\textrm{pipe}$ denotes the number of pipeline stages (five in this context). This is because the execution of two consecutive gates presents data dependencies. Consequently, for circuits with qubits count lower than $\ceil{\log_2{(N_\textrm{pipe})+2}}$, stalls must be inserted to ensure the correctness of the results.
For saving area, it was decided to compute the update of at least $2^{N_{q_\textrm{min}}}$ couples also for smaller circuits but not to store the outcomes exceeding $2^{N_q}$. This approach eliminates the need to instantiate a dedicated unit to manage stalls while maintaining the same time penalty.\\
The pipeline can be introduced by standardizing the execution of the supported gates and recognizing that all can be implemented as: 
\vspace{-3pt}
\begin{align}
\begin{split}
     c_{i_{\textrm{out}}} &= \alpha \sin{(\theta)} + \beta\cos{(\theta)} +  i( \gamma \sin{(\theta)} + \delta \cos{(\theta)} )\\
    c_{j_{\textrm{out}}} &= \epsilon \sin{ (\theta)} + \zeta \cos{(\theta)}  + i( \eta \sin{(\theta)} + \iota \cos{(\theta)} ) \, ,
    \end{split}
    \vspace{-10pt}
    \label{eq:computationMechanism}
\end{align}
where $\alpha$, $\beta$, $\gamma$, $\delta$, $\epsilon$, $\zeta$, $\eta$ and $\iota$ are properly chosen depending on the gate for selecting real or imaginary parts of the probability amplitudes in input, $c_{i_{\textrm{out}}}$ and  $c_{j_{\textrm{out}}}$ are the couple of probability amplitudes associated with $i^{\textrm{th}}$ and $j^{\textrm{th}}$ basis states in the output state vector and $\theta$ --- immediate field of the instruction --- is the parametric angle in the rotational gate and a gate-dependent fixed angle in the others. The sine and cosine values are exploited to change the sign or delete a factor, thanks to trigonometric properties. 
Therefore, the datapath comprises four computing units --- one for the real, one for the imaginary part of each probability amplitude ---, containing two multipliers and an adder.  Sine and cosine are computed by the  \textbf{TU}. \\
The \textbf{TU} implements the architecture presented in \cite{de2014fixed}, which proposed an efficient approach based on the exploitation of look-up tables (LUTs) and the Taylor series. This solution should help the target application achieve a \textbf{better balance between area and accuracy}, compared to a solution based on the COordinate Rotation DIgital Computer (CORDIC) algorithm. Although CORDIC can be fully unrolled, the considered approach involves fewer processing elements. All the emulator blocks are synchronized and managed by the \textbf{QECU}. \\
The interaction with the external is implemented through \textbf{asynchronous First-In-First-Out} (\textbf{FIFO}) buffers (platform independent) and a \textbf{communication interface unit} (platform dependent),  both coordinated by a communication control unit. The two buffers, one for transmission of the state vector and the other for the reception of the instruction to execute, handle the \textbf{clock domain crossing} to avoid metastability issues. The communication interface unit is the only thing that should be modified varying the platform. In this context, it implements the AMBA 4 AXI4-Stream communication ARM protocol, permitting the exploitation of the Direct Memory Access (DMA) mechanism. 

{\renewcommand{\arraystretch}{1.5}
\setlength{\tabcolsep}{1.2pt}
\begin{table}[t]
\caption{Comparison between AMARETTO synthesis results and the current literature. \vspace{-10pt} }
\begin{center}
\resizebox{1\columnwidth}{!}{
\begin{tabular}{?c?c?c?c?c?c?c?c?}
\noalign{\hrule height 1.5pt}
\textbf{Emulator} & \textbf{AMARETTO} & \cite{pilch2019fpga} & \cite{mahmud2018scalable} & \cite{mahmud2020efficient} & \cite{fujishima200316} & \cite{conceiccao2015efficient} \\
\noalign{\hrule height 1.5pt}
\textbf{$N_{\textrm{qubit}}$} & 16 & 2 & 4 & 32  & 16 & 9 \\
\hline
\makecell{\textbf{Device}} &  \makecell{AMD Kria\\ KV260} & \makecell{Intel \\Cyclone V} & \makecell{Intel \\Arria 10} & \makecell{Intel \\Arria 10} &  \makecell{Intel(APEX) \\20KE1500} & \makecell{Intel\\ Stratix} \\
\hline
\textbf{BRAM} & 2.62 MB & - & 32.08 MB & 32 GB (ext.) & - & - \\
\hline
\makecell{\textbf{Logic} \\ \textbf{Utilization}} &  \makecell{7751/117120\\ CLB} & \makecell{8000\\ ALMs} & \makecell{374021\\ ALMs} & \makecell{56219\\ ALMs} & \makecell{1500000*\\ Gates} & \makecell{4019 \\LC} \\
\hline
\textbf{DSP} &  11/1248 & - & 1364 & 49 & - & - \\
\hline
\textbf{Precision} &  20-bit fixed & 10-bit fixed  & 32-bit float & 64-bit float & - & 18-bit fixed \\
\hline
\textbf{$f_\textrm{clk}$ [MHz]} & \SI{100}{\mega \hertz}  & - & \SI{233}{\mega \hertz}& \SI{233}{\mega \hertz} & \SI{60}{\mega \hertz}  & - \\
\noalign{\hrule height 1.5pt}
\end{tabular}} 
\end{center}
\label{tab:TableComparisons}
\vspace{-40pt}
\end{table}

\section{Results} \label{sec:results}
\vspace{-3pt}
\noindent The architecture synthesis on the \textbf{AMD Kria KV260 SoM} using \textbf{Vivado 2023.1} achieved a maximum of \textbf{sixteen qubits}. 
\textbf{RAM availability} emerged as the \textbf{bottleneck}, reaching 100\% utilization. This aligns with expectations as memory dominates due to its $\mathcal{O}(N_\textrm{bit} 2^{N_q})$ scaling, where $N_\textrm{bit}$ is the number of bits exploited for numerical representation.\\
The obtained synthesis results are compared with the current literature in Table \ref{tab:TableComparisons}. Making direct comparisons poses challenges due to differences among various architectures in the different target FPGAs, produced by different companies. Furthermore, the reported information in these articles is often incomplete, and the synthesis of many of these architectures depends on the quantum circuit, differently from AMARETTO.  Indeed, its advantage lies in \textbf{not requiring re-synthesis} for executing new circuits, unlike other architectures. Additionally, its \textbf{extensive gate support} allows it to handle all applications below the platform's maximum capacity efficiently.\\
However, some observations can still be made. Although the board considered in this work is relatively small, \cite{mahmud2020efficient} is the only architecture achieving a higher number of emulable qubits leveraging external memory.  Nevertheless, \cite{mahmud2020efficient} architecture demands relatively more memory in proportion to the number of qubits with respect to AMARETTO, as it employs more bits for number representation, storing both the state vector and the gate matrices.\\
Additionally, the relative logic occupation of our architecture is the lowest among the compared designs. Furthermore, despite operating at a lower frequency than \cite{ mahmud2018scalable,mahmud2020efficient}, AMARETTO is expected to be faster since a single gate execution requires $\mathcal{O}(N)$ clock periods instead of $\mathcal{O}(N^2)$, where $N$ is the length of the state vector, i.e. $2^{N_q}$. \\
Functional verification involved about \textbf{fifty quantum circuits} in OpenQASM 2.0, compared with \href{https://qiskit.github.io/qiskit-aer/stubs/qiskit_aer.StatevectorSimulator.html}{Qiskit's state vector simulator} using \textbf{Great-circle distance (GCD)}, thus considering state vector elements in polar coordinate, i.e. as points on a sphere, and their spheric distance estimates the divergence between the two results. The GCD accentuates the differences between complex numbers, thus guaranteeing a more reliable functional validation of the architecture. GCD consistently remained below 0.05, meeting the study's acceptability threshold.
\begin{figure}[h]
    \centering \vspace{-10pt}
    \includegraphics[width=0.7\columnwidth]{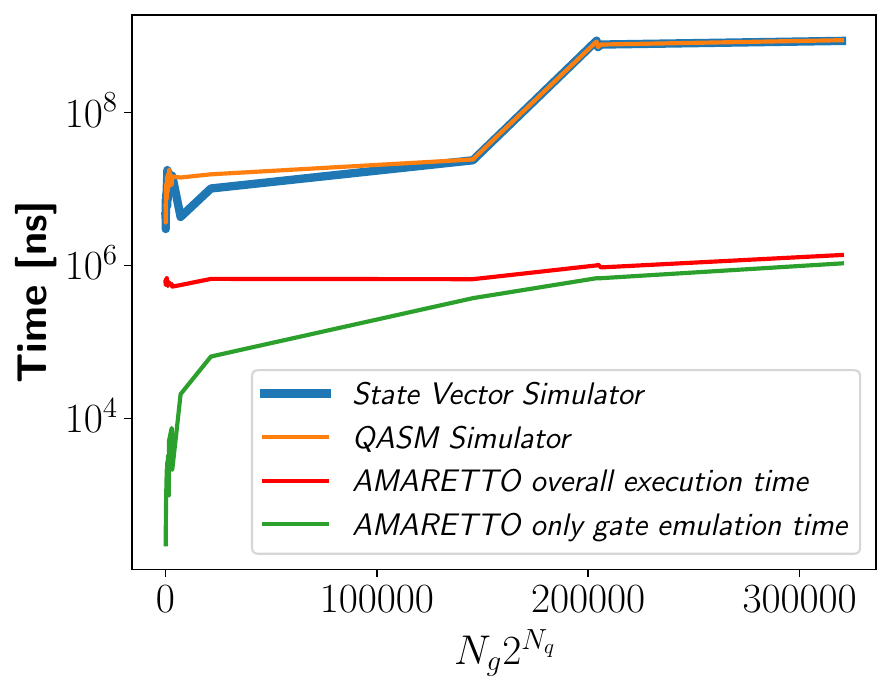} \vspace{-8pt}
    \caption{Comparison of execution time between Qiskit simulators and AMARETTO, showing hardware emulation's significant advantage over software.
    \vspace{-20pt}}
    \label{fig:Time}
\end{figure}\\
Figure \ref{fig:Time} compares execution times of Qiskit simulators (\href{https://qiskit.github.io/qiskit-aer/stubs/qiskit_aer.QasmSimulator.html}{Qasm} and \href{https://qiskit.github.io/qiskit-aer/stubs/qiskit_aer.StatevectorSimulator.html}{State Vector}) --- on a single-process Intel(R) Xeon(R) Gold 6134 CPU @ 3.20 GHz opta-core, Model 85, with a memory of about 103 GB \cite{processorservervlsi} --- and AMARETTO, demonstrating hardware emulation's \textbf{orders of magnitude lower} time requirements, particularly evident with larger quantum circuits. The execution time scales as $\biggl(2^{\max(N_q, N_{q_\textrm{min}})-1}\frac{N_g(2-\alpha)}{2}+(N_{\textrm{pipe}}-1)\biggr)T_{\textrm{clock}}$, where $\alpha$ is the percentage of controlled gates and $T_{\textrm{clock}}$ the clock period. Therefore, it scales linearly with $2^{N_q}N_g$.

\vspace{-3pt}
\section{Conclusions}\label{sec:conclusions}
\vspace{-3pt}
This article introduces AMARETTO, a specialized architecture for quantum computing emulation on low-tier FPGAs, supporting Clifford+T and rotational gate sets. A comparative analysis shows AMARETTO's execution time is about two orders of magnitude faster than the Qiskit state vector simulator. Emulating sixteen qubits on a AMD Kria KV260 SoM, AMARETTO matches the qubit capacity of other works using a smaller and cheaper FPGA. This promising solution addresses challenges in validating new quantum algorithms, potentially advancing quantum computing application development.


 \vspace{-5pt}
\bibliographystyle{ieeetr}
\bibliography{acmart}

\end{document}